\documentclass[aps,pre,twocolumn,showpacs,amsmath,amssymb,floatfix]{revtex4-1}
\usepackage[utf8x]{inputenc} 
\usepackage{graphicx}
\usepackage{multirow}
\usepackage{rotating}
\usepackage{amssymb,color}

\begin{document}

\title{Wang-Landau sampling: Improving accuracy}

\author{A. A. Caparica$^{\dagger}$ and A. G. Cunha-Netto$^{\dagger\dagger}$}

\affiliation{$\dagger$ Instituto de F\'{\i}sica, Universidade Federal de
Goi\'{a}s.  C.P. 131, CEP 74001-970, Goi\^{a}nia, GO, Brazil \\
$\dagger\dagger$ Departamento de F\'{\i}sica, Instituto de Ci\^{e}ncias Exatas,
and National Institute of Science and Technology for Complex Systems,
Universidade Federal de Minas Gerais, C.P.702, 30123-970 Belo Horizonte, Minas
Gerais, Brazil}

\begin{abstract}
In this work we investigate the behavior of the microcanonical and canonical averages
of the two-dimensional Ising model during the Wang-Landau simulation. The simulations were
carried out using conventional Wang-Landau sampling and the $1/t$ scheme. Our findings
reveal that the microcanonical average should not be accumulated during the initial modification factors \textit{f}
and outline a criterion to define this limit. We show that updating the density of states only after every $L^2$
spin-flip trials leads to a much better precision. We present a mechanism to determine for the given model up to
what final modification factor the simulations should be carried out.  Altogether these
small adjustments lead to an improved procedure for simulations with much more reliable results.
We compare our results with $1/t$ simulations. We also present an application of the procedure to a
self-avoiding homopolymer.
\end{abstract}
\maketitle

\section{Introduction}

In recent years Wang-Landau sampling (\textit{WLS})\cite{wls1,wls2} has been applied to many systems and has become a
well-established Monte Carlo algorithm. The heuristic idea of the method is based on the fact that if one performs
a random walk in energy space with a probability proportional to the reciprocal of the density of states,
a flat histogram is generated for the energy distribution. Since the density of states produces huge numbers,
instead of estimating $g(E)$, the simulation is performed for $S(E)\equiv\ln g(E)$, and a histogram $H(E)$ is accumulated
during the simulations to control the frequency of visits to the energy levels. At the beginning
of the simulation we set $S(E)=0$ for all energy levels. The random walk in the energy space runs through all energy levels from $E_{min}$ to
$E_{max}$ with a probability

\begin{equation}\label{prob}
p(E\rightarrow E')=\min(\exp{[S(E)-S(E')]},1) ,
\end{equation}
where $E$ and $E'$ are the energies of the current and the new possible configurations. Whenever a configuration
is accepted we update $H(E')\rightarrow H(E')+1$ and $S(E')\rightarrow S(E')+F$,
where $F=\ln f$,($f$ is the so-called modification factor). The initial modification factor is taken as
$f=f_{0}=e=2.71828...$. If the trial configuration is not accepted, then the current $H(E)$ and $S(E)$ are updated
again. The flatness of the histogram is checked after a number of Monte Carlo (MC) steps
and usually the histogram is considered flat if $H(E)>0.8\langle H \rangle$, for all energies, where
$\langle H \rangle$ is an average over the energies. If the flatness condition is fulfilled we update the modification
factor to a finer one by setting $f_{i+1}=\sqrt{f_{i}}$ and reset the histogram $H(E)=0$. Simulations are in general halted when $\ln f\sim10^{-8}$. Having in hand the density of states, one can calculate the canonical average of any thermodynamic variable as

\begin{equation}\label{mean}
\langle X\rangle_T=\dfrac{\sum_E \langle X\rangle_E g(E) e^{-\beta E}}{\sum_E g(E) e^{-\beta E}} ,
\end{equation}
where $\langle X\rangle_E$ is the microcanonical average accumulated during the simulations and $\beta=1/k_BT$,
$k_B$ is the Boltzmann constant and $T$ is the temperature. One of the
interesting features of the method is that it can also access some quantities, such as the free energy and
entropy, which are not directly available from conventional Monte Carlo simulations.

As described above, the convergence of the method depends on both the flatness criterion and the final \textit{f}
when the simulation is interrupted, but the best choice of each is not obvious for each model to be studied.

Recently some authors have asserted that although achieving a flat histogram is the initial
motivation of the \textit{WLS}, the flatness is not a necessary criterion
to reach convergence \cite{zhou,belardinelli,belardinelli2,belardinelli3}.
They argue that in conventional \textit{WLS} the error saturates to a constant, while if $\ln f$
decreases as $1/t$, where $t$ is a normalized Monte Carlo time, the
error would decrease monotonically as well. The $1/t$ algorithm is divided into two steps,
initially the conventional \textit{WLS} is followed, starting from $S(E)=0$ and then constructing $S(E)$
using a histogram updated in every new accepted configuration. $S(E)$  is updated as in the conventional \textit{WLS},
$S(E)=S(E)+F_i$, with the initial value $F_0=1$.
After a number of moves (e.g. $1000$ MC sweeps), we check $H(E)$ to verify whether all the levels were visited
by the random walker at least once and then update $F_i=F_i/2$ and reset $H(E)=0$.
(The flatness criterion is not required, even in this first stage.) Simulation is performed
while $F_i\geq 1/t=N/j$, where $j$ is the number of trial moves
and $N$ is the number of energy levels. In the remainder of the simulation $F_i$ is updated every new
configuration as $F_i=1/t$ up to a final chosen precision $F_{final}$.

The efficiency, convergence and limitations of the \textit{WLS} has been quantitatively studied \cite{troyer,bhatt}.
In the present work we perform a practical, computational study on the convergence and the accuracy of the method.

In this paper we investigate the behavior of the maxima of the specific heat
\begin{equation}\label{cv}
 C=\langle(E-\langle E\rangle)^2\rangle/T^2,
\end{equation}
and the susceptibility
\begin{equation}\label{ki}
 \chi=L^2\langle(m-\langle |m|\rangle)^2\rangle/T,
\end{equation}
where $E$ is the energy of the configurations and $m$ is the corresponding magnetization per spin during the conventional
\textit{WLS} and the $1/t$ algorithm simulations for the Ising model
on a square lattice \cite{plichske}. We observe (as in \cite{belardinelli,belardinelli2,belardinelli3,swetnam})
that a considerable part of the conventional Wang-Landau simulation
is not very useful because the error saturates. We propose some strategies to improve the efficiency of \textit{WLS}
and compare our results with exact calculations \cite{beale}.
Our findings lead to a new way of performing the \textit{WLS} simulations.

\maketitle

\section{A new procedure for simulations}

\begin{figure}[!ht]\centering
\begin{center}
 \includegraphics[width=.7\linewidth,angle=-90]{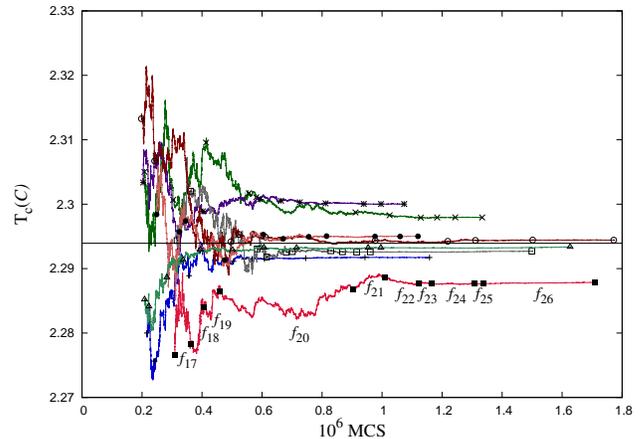}
\end{center}
\caption{Evolution of the temperature of the extremum of the specific heat during the \textit{WLS},
beginning from $f_{17}$, for eight independent runs using the $80\%$-flatness criterion. The dots show where the modification factors were updated
and the straight line is the result obtained using the exact data from Ref. \cite{beale}.}
\label{medcan_cv_conv}
\end{figure}

In order to test how far the simulations should go, during the \textit{WLS}, beginning from $f_{17}$, we calculate the specific heat and the susceptibility defined in Eqs.\eqref{cv} and \eqref{ki}, as well as the energy and the magnetization at some fixed temperatures. We use the current $g(E)$ and from this time on this mean values are updated whenever we check the flatness of the histogram. Fig.\ref{medcan_cv_conv} shows the evolution
of the temperature of the maximum of the specific heat calculated for $L=32$ for eight
independent runs as a function of the Monte Carlo sweeps (MCS) \cite{mcs} and
compare these results with the value obtained using the exact data of Ref. \cite{beale} ($T_c(L=32)=2.29392979$). The dots
label the MCS when the modification factor was updated, the leftmost in each run corresponding to $f_{17}$.
One can see that around $f_{23}=1.1921\times10^{-7}$ all the curves become stabilized in values displaced close to the exact value.
Any further computational effort for $\ln f<\ln f_{23} $ does not lead to a better convergence.

\begin{figure}[!ht]\centering
\begin{center}
 \includegraphics[width=.7\linewidth,angle=-90]{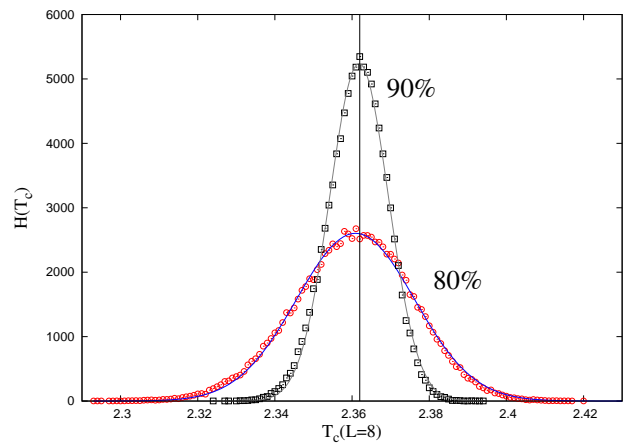}
\end{center}
\caption{Histograms of the locations of the peak of the specific heat for the 2D Ising model
during the \textit{WLS}, using the $80\%$- and $90\%$-flatness criterions, each for $100,000$ independent
runs, along with their best-fit Gaussians. The central line corresponds to the exact temperature of the
maximum of the specific heag obtained with data from Ref.\cite{beale}.}
\label{histograms_80_90}
\end{figure}

\begin{figure}[!ht]\centering
\begin{center}
 \includegraphics[width=.7\linewidth,angle=-90]{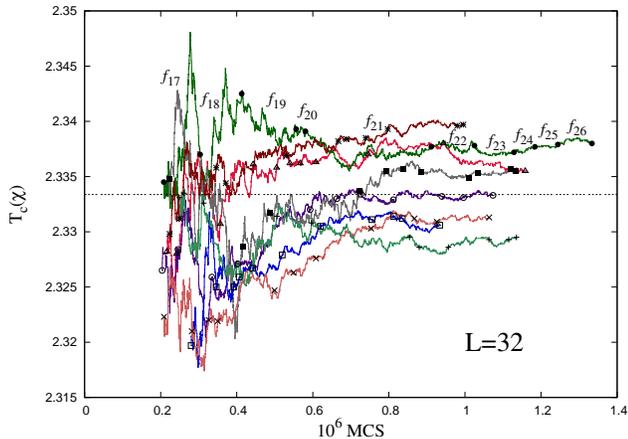}
\end{center}
\caption{Evolution of the temperature of the extremum of the susceptibility during the \textit{WLS}, beginning from $f_{17}$, for eight independent runs using the $80\%$-flatness criterion. The dots show where
the modification factors were updated and the straight line is the result obtained
using the exact data from Ref. \cite{beale}.}
\label{medcan_khi_conv}
\end{figure}

In order to investigate how these results are displaced around the exact value, we performed $100,000$ independent runs of \textit{WLS} for $L=8$ using the $80\%$ and the $90\%$ flatness criterions and built up histograms using bins of width 0.001.
In Fig.\ref{histograms_80_90} we show that the histograms form nice Gaussians centered close, but not precisely in the exact
value. In Fig.\ref{medcan_khi_conv} we show the same evolution for the temperature of the maximum of the susceptibility.
One can see that in this case the curves do not flow to steady values.

\begin{figure}[!ht]\centering
\begin{center}
 \includegraphics[width=.7\linewidth,angle=-90]{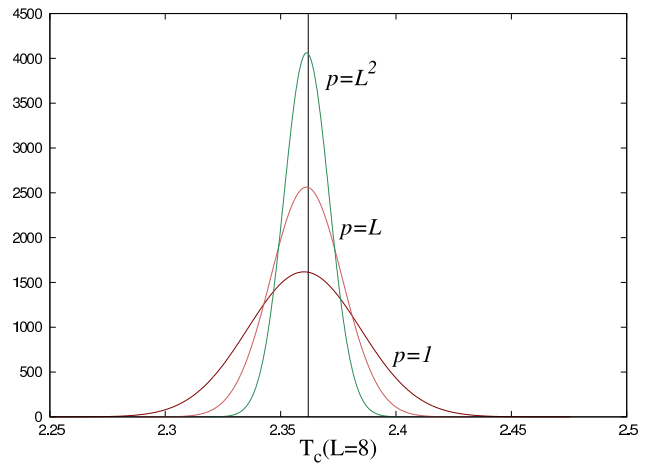}
\end{center}
\caption{Best-fit Gaussians for the histograms of the temperatures of the peak of the specific heat for 2D Ising model
during the \textit{WLS} up to $\ln f=10^{-4}$, using the $80\%$-flatness criterion, each for $100,000$ independent
runs with the density of states being updated every $p$ spin-flip trials. The central line
corresponds to the exact temperature obtained with data from Ref.\cite{beale}}.
\label{gaussians_p}
\end{figure}

A strategy to improve the precision of the \textit{WLS} is to update the density of states periodically (i.e.,
after every $p$ trial configurations), instead of updating $S(E)$ every spin-flip trial. In order to
investigate how this change affects the final result, we performed $100,000$ independent runs ($L=8$)
using the $80\%$ flatness criterion and constructed again histograms of the locations
of the peak of the specific heat. We tested the \textit{WLS} with different values for $p$. Fig. \ref{gaussians_p} shows the Gaussian best-fits for $p=1$ (conventional \textit{WLS}), $p=L$ and $p=L^2$. The vertical line indicates the exact value
using Ref. \cite{beale}. One can see that the higher the values of $p$, the narrower the Gaussian curves. Defining the
relative error $\varepsilon(X)$ for any quantity $X$ by

\begin{equation}\label{error}
\varepsilon(X)=\frac{|X_{sim}-X_{exact}|}{X_{exact}} ,
\end{equation}
we obtain the relative errors of the simulated mean values with respect to the result using Ref. \cite{beale} for $p=1$, $L$ and $L^2$ as $0.00043$, $0.00019$ and $0.00013$, respectively. Therefore, we see that updating the density of states only after
$L^2$ trial moves leads to more accurate results \cite{blume,blume_l2}. In other words, adopting the Monte Carlo step ($L^2$
trial moves), as it is conventional in the Metropolis algorithm, is also convenient in the Wang-Landau algorithm.

\begin{figure}[!ht]\centering
\begin{center}
 \includegraphics[width=.7\linewidth,angle=-90]{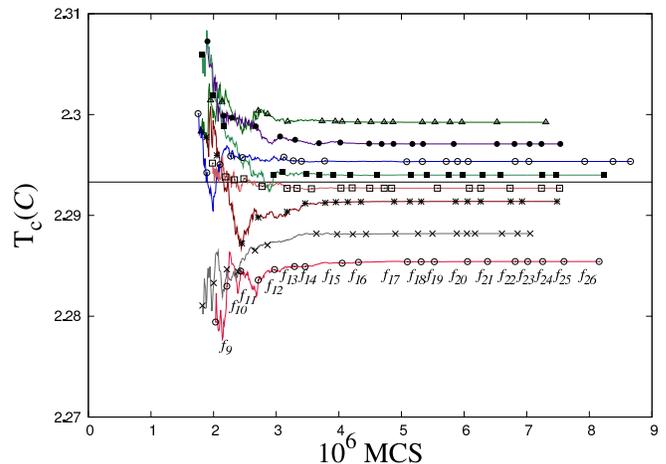}
\end{center}
\caption{Evolution of the temperature of the extremum of the specific heat during the \textit{WLS},
beginning from $f_{9}$, for eight independent runs. The density of states were updated after every
$L^2$ trial moves and the flatness criterion was $80\%$. The dots show where the modification factor
was updated and the straight line is the result obtained using the exact data from Ref. \cite{beale}.}
\label{medcan_cv_l2}
\end{figure}

In Fig.\ref{medcan_cv_l2} we show the evolution of the location of the maximum of the heat capacity during $WLS$ in
which the density of states was updated only after every $L^2$ spin-flip trials, beginning from $f_9$. We see that
now the curves flow to steady values around $\ln f=\ln f_{13}=1.2208\times10^{⁻4}$ and simulations with higher orders of the modification factor are unnecessary. One can see that, if on one hand the simulations adopting the Monte Carlo step becomes
longer, on the other hand, the canonical averages converge to constant values much earlier.

\begin{figure}[!ht]\centering
\begin{center}
\centerline{\includegraphics[width=\linewidth]{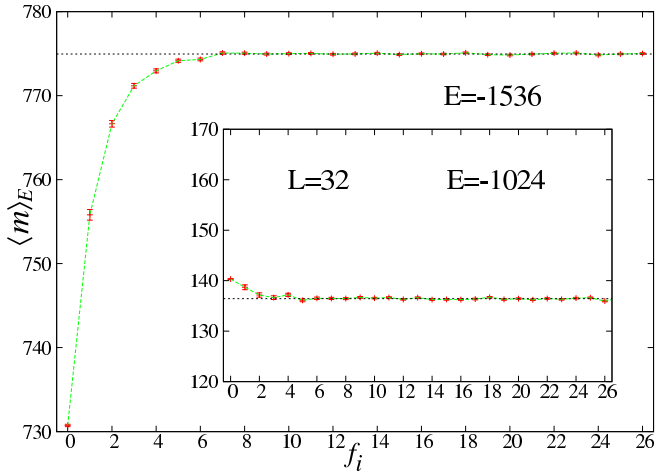}}
\end{center}
\caption{Evolution of the microcanonical average of the magnetization for 2D Ising model for $L=32$ at $E=-1024$ and $-1536$
during the simulations over $1000$ independent runs for each flatness stage.}
\label{micro}
\end{figure}

We now turn our attention to another important detail. What is the behavior of the microcanonical averages \hspace{1mm}
$\langle M\rangle_E$ \hspace{1mm} and \hspace{1mm} $\langle M^2\rangle_E$ \hspace{1mm} during the sampling process?
We have also evaluated the microcanonical averages during the simulations. In order to estimate the mean value of the
magnetization during each flatness stage we carried out $1000$ independent runs and calculated $\langle M\rangle_E$ for
each $f_i$ with $i=0,1,2,...,26$. In Fig.\ref{micro} we show these results for two energy levels and see that they flow
to relatively stable values around $f_7$. We therefore conclude that the microcanonical averages should not be accumulated
before $\ln f=\ln f_7=7.843\times10^{-3}$.

\begin{figure}[!ht]\centering
\begin{center}
 \includegraphics[width=.7\linewidth,angle=-90]{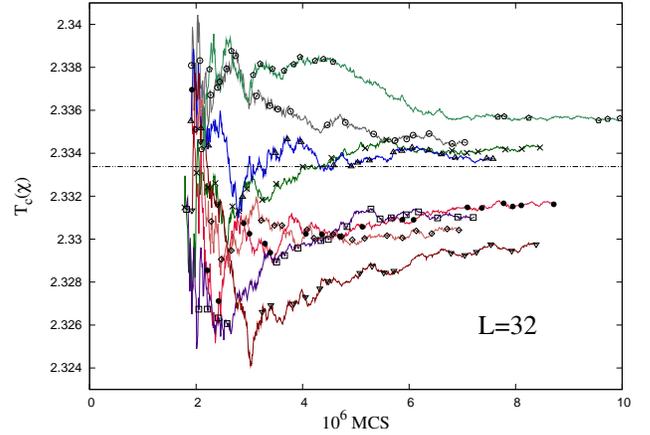}
\end{center}
\caption{Evolution of the temperature of the extremum of the susceptibility during the \textit{WLS},
beginning from $f_{9}$, for eight independent runs. The density of states was updated after every
$L^2$ trial moves and the flatness criterion was $80\%$. The dots show where the modification factor
was updated and the straight line is the result obtained using the exact data from Ref. \cite{beale}
with the microcanonical average accumulated
from $\ln f=\ln f_7$.}
\label{medcan_khi_l2_comum}
\end{figure}

\begin{figure}[!ht]\centering
\begin{center}
 \includegraphics[width=.7\linewidth,angle=-90]{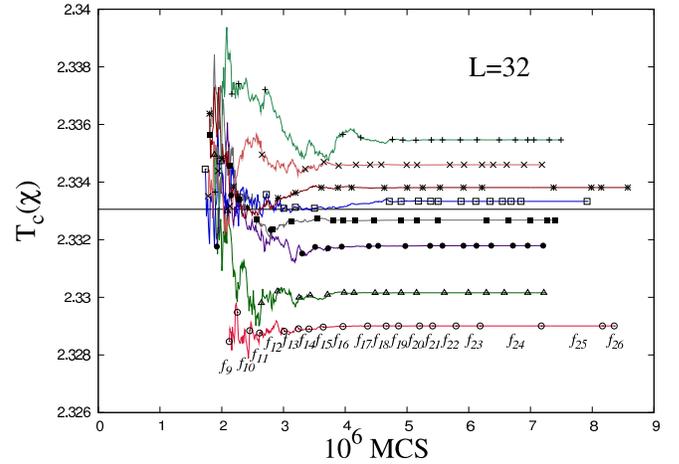}
\end{center}
\caption{Evolution of the temperature of the extremum of the susceptibility during the \textit{WLS},
beginning from $f_{9}$, for eight independent runs, using a common microcanonical average in $24$
independent runs. The density of states was updated after every $L^2$ trial moves and the flatness
criterion was $80\%$. The dots show where the modification factor was updated and the straight line
is the result obtained using the exact data from Ref. \cite{beale} with the microcanonical average accumulated
from $\ln f=\ln f_7$.}
\label{medcan_khi_l2_restrito}
\end{figure}

In Fig.\ref{medcan_khi_l2_comum} we show the evolution of the maximum of the susceptibility during the simulations beginning
from $f_9$, updating the density of states after every $L^2$ spin-flip trials and accumulating the microcanonical averages
only for $\ln f \leq\ln f_7$. We observe that even for $\ln f=\ln f_{26} \approx10^{-8}$ we do not obtain stable values like those of
Fig.\ref{medcan_cv_l2}. However, if one takes the mean value of the microcanonical averages in $24$ independent runs and uses
this result for calculating the canonical averages during the simulations the averages do flow to stable values, as shown in
Fig.\ref{medcan_khi_l2_restrito}. This result shows that even for quantities that involve the magnetization the simulations
can be carried out only up to $\ln f=\ln f_{13}$.

The evolution of the canonical averages of the energy and the magnetization at a given temperature yields evidently
patterns similar to those of Fig.\ref{medcan_cv_l2} and Fig.\ref{medcan_khi_l2_restrito} with the same limit modification
factors.

We would like to stress that all modification factors defined above using the canonical e microcanonical averages during
the simulations apply to the 2D Ising model and it is important to perform studies similar to those of
Figures \ref{medcan_cv_l2}, \ref{micro} and \ref{medcan_khi_l2_restrito} before adopting this new procedure to other models to be sure on where to halt the simulations and where to begin accumulating the microcanonical averages.

In view of the above observations, we propose the following new procedure for simulations:

\begin{itemize}

\item {Instead of updating the density of states after every spin-flip, we ought to update it after each Monte Carlo sweep;}

\item {\textit{WLS} should be carried out only up to a $\ln f=\ln f_{final}$ defined by the canonical
averages during the simulations;}

\item {The microcanonical averages should not be accumulated before $\ln f\leq\ln f_{micro}$ defined by the microcanonical
averages during the simulation.}

\end{itemize}

\maketitle

\section{Comparison with $1/t$ simulations}

\begin{figure}[!ht]\centering
\begin{center}
 \includegraphics[width=.7\linewidth,angle=-90]{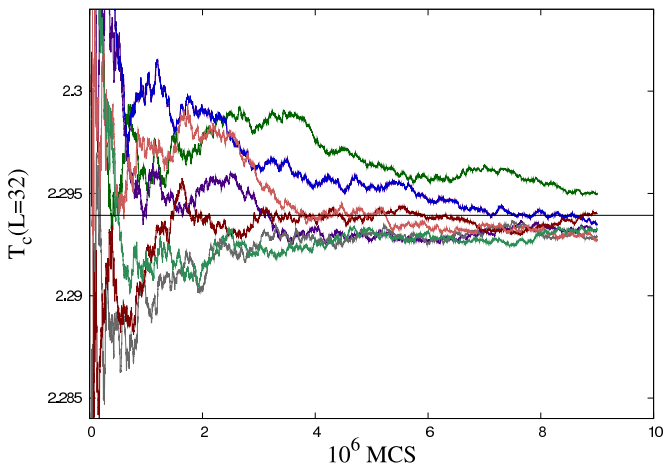}
\end{center}
\caption{Evolution of the temperature of the extremum of the specific heat during the \textit{1/t} simulations
for eight independent runs beginning from the second stage. The straight line is the result obtained using the exact data from Ref. \cite{beale}. Simulations were halted when the CPU time matched up the mean time of \textit{WLS}.}
\label{medcan_1_t}
\end{figure}

Fig.\ref{medcan_1_t} shows the evolution of the maxima of the specific heat during the simulations using the
$1/t$ scheme, beginning from the second stage and halting the simulations when the CPU time matched up the
mean time of the simulations of Fig.\ref{medcan_cv_l2}.

\begin{figure}[!ht]\centering
\begin{center}
 \includegraphics[width=.7\linewidth,angle=-90]{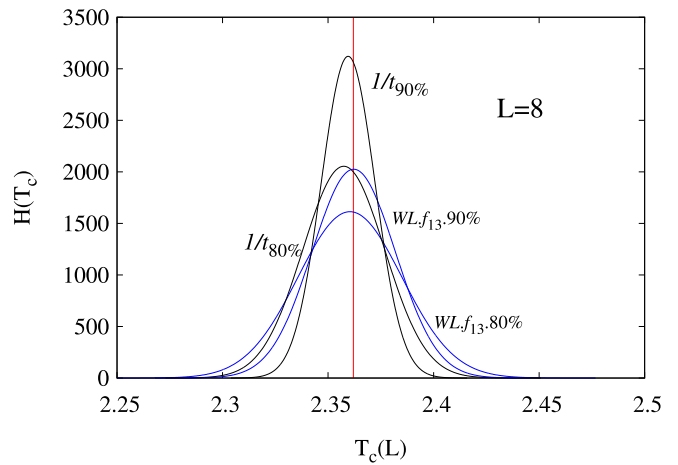}
\end{center}
\caption{Best-fit Gaussians for the histograms of the temperatures of the peak of the specific heat for 2D Ising model
during the \textit{WLS} up to $\ln f=10^{-4}$, using the $80\%$- and $90\%$-flatness criterions, each for $100,000$ independent
runs. The $1/t$ simulations were carried out within the same CPU time. The central line corresponds to the exact temperature
obtained with data from Ref.\cite{beale}.}
\label{histograms}
\end{figure}

In order to compare these results we performed $100,000$ independent runs of \textit{WLS} for $L=8$ up to
$\ln f=\ln f_{13}$ using $80\%$ and $90\%$ flatness criterions ($WL.f_{13}.80\%$ and $WL.f_{13}.90\%$)
and built up the histograms. Next we carried out similar simulations using the $1/t$ algorithm, halting the simulation
when the CPU time matched up those of $WL.f_{13}$ ($1/t_{80\%}$ and $1/t_{90\%}$). In Fig.\ref{histograms} we show the
best-fit Gaussians of the histograms. One can see that they are not really centered around the exact value. The
relative errors of the simulated mean values with respect to the result using Ref. \cite{beale} yield $0.00041$
and $0.00036$, respectively, for $WL.f_{13}.80\%$ and $WL.f_{13}.90\%$, and $0.0017$ and $0.00081$ for $1/t_{80\%}$
and $1/t_{90\%}$, with final $F_k$ reaching $5.1\times10^{-7}$ and $2.4\times10^{-7}$. We see that although the
widths of the  $1/t$-curves are smaller, their centers are farther apart from the exact value than those
of $WLS$ revealing a biased estimation effect in the $1/t$ method.

\section{Finite-size scaling}

According to finite-size scaling theory \cite{fisher1,fisher2,barber} from the definition of the
free energy one can obtain the zero field scaling expressions for the magnetization and the susceptibility, respectively by

\begin{equation}\label{exp1}
 m\approx L^{-\beta/\nu}\mathcal{M}(tL^{1/\nu}),
\end{equation}

\begin{equation}\label{exp2}
\chi \approx L^{\gamma/\nu}\mathcal{X}(tL^{1/\nu}).
\end{equation}

\noindent We see that the locations of the maxima of these functions scale asymptotically as
\begin{equation}\label{tc}
 T_c(L) \approx T_c+a_qL^{-1/\nu},
\end{equation}
where $a_q$ is a quantity-dependent constant, allowing then the determination of $T_c$.

In order to compare the efficiency of the conventional \textit{WLS}, the \textit{1/t}-scheme and our procedure, we performed
simulations with $L = 32, 36, 40, 44,  48, 52, 56, 64, 72$ and $80$ taking $N = 24, 24, 20, 20, 20, 16, 16, 16, 12$
and $12$ independent runs for each size, respectively.

\begin{figure}[!ht]
\centerline{\includegraphics[width=.9\linewidth]{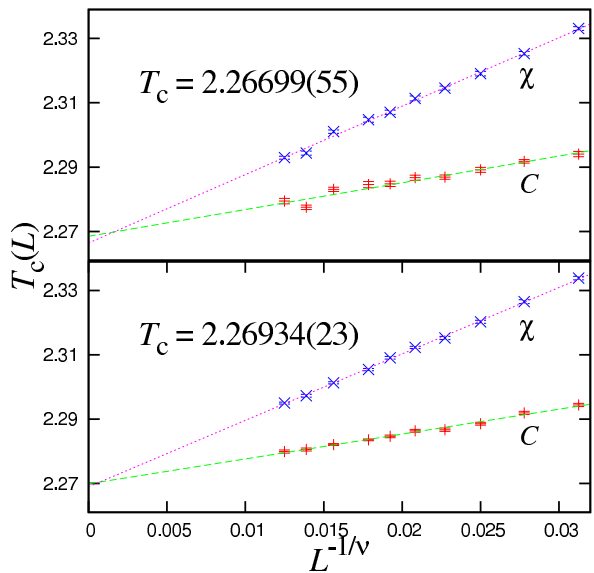}}
\caption{Size dependence of the locations of the extrema in the specific heat and the susceptibility for conventional
\textit{WLS} (top) and using our procedure (bottom) assuming $\nu = 1$.}
\label{tc}
\end{figure}
\vspace*{.3cm}
\begin{table}[!hb]
  \centering
  \begin{tabular}{c c c c c}
  \hline\hline
  Case &   $T_c$  & $\beta$ &  $\gamma$ & CPU time \\ [0.5ex]
  \hline
  Exact & 2.2691853... & 0.125 & 1.75& \\
  \hline
        \multicolumn{5}{c}{$1/t$}       \\
  \hline
 $1.10^{-6}$ & 2.2621(11)  & 0.197(14)  & 1.943(35)&  0.15\\
 $5.10^{-7}$ & 2.2642(11)  &0.1479(84)  & 1.846(18) &  0.30\\
 $1.10^{-7}$ & 2.26848(35) & 0.1297(31) & 1.7833(46)&  1.51\\
 $5.10^{-8}$ & 2.26904(25) & 0.1259(21) & 1.7708(23)&  3.03\\
 $1.10^{-8}$ & 2.26944(11) & 0.12647(94) & 1.7616(17)& 15.13\\
  \hline
       \multicolumn{5}{c}{Conventional \textit{WLS}}    \\
  \hline
    $80\%$  & 2.26699(55) & 0.1295(45) & 1.7812(63) & 1.00\\
    $90\%$  & 2.26829(33) & 0.1386(51) & 1.7899(87) & 1.75\\ [1ex]
  \hline
        \multicolumn{5}{c}{Our procedure} \\
  \hline
    $80\%$  & 2.26934(23) & 0.1270(16) & 1.7631(27) & 9.78\\
    $90\%$  & 2.26916(12) & 0.12494(68) & 1.7555(32) & 22.21\\ [1ex]
 \hline\hline
  \end{tabular}
  \caption{Finite size scaling results for the critical temperature and the critical exponents $\beta$
and $\gamma$. The CPU times are expressed in terms of the time spent by the conventional \textit{WLS} with $80\%$-flatness.}
  \label{table}
  \end{table}
\begin{figure}[!ht]
\centerline{\includegraphics[width=.9\linewidth]{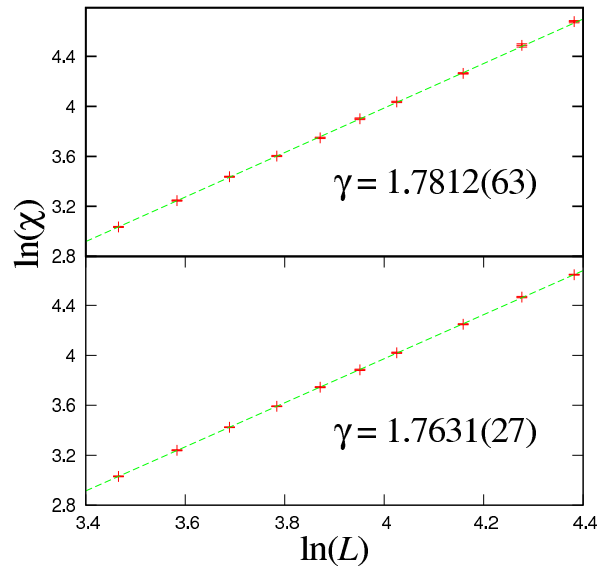}}

\caption{Log-log plot of size dependence of the finite-lattice susceptibility at $T_c(L)$ with $80\%$ flatness criterion
for conventional \textit{WLS} (top) and using our procedure (bottom).}
\label{gamma}
\end{figure}
\begin{figure}[!ht]
\centerline{\includegraphics[width=.9\linewidth]{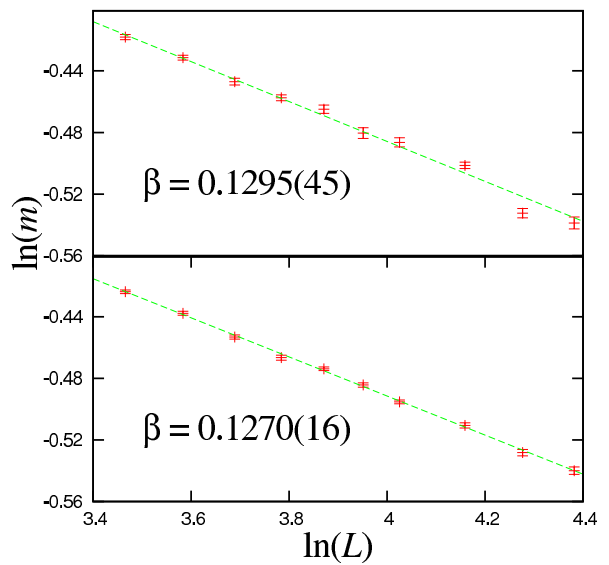}}

\caption{Log-log plot of size dependence of the finite-lattice magnetization with $80\%$ flatness criterion
for conventional \textit{WLS} at $T_c = 2.26699$ (top) and using our procedure at $T_c = 2.26934$ (bottom).}
\label{beta}
\end{figure}

Using these scaling functions we estimated the critical temperature and the critical
exponents $\beta$ and $\gamma$. Taking a microcanonical average including all independent runs was
important to reveal in Fig.\ref{medcan_khi_l2_restrito} that for quantities that involve the magnetization
the simulations can also be carried out only up to $\ln f=\ln f_{13}$, but such a procedure does
not lead to better results for the estimation of the canonical averages.

Assuming $\nu=1$ we can use Eq.\eqref{tc} to determine $T_c$
as the extrapolation to $L\rightarrow\infty$ ($L^{-1/\nu}=0$) of the linear fits given by the locations
of the maxima of the specific heat and the susceptibility defined by Eqs.\eqref{cv}-\eqref{ki}.
In Fig.\ref{tc} we show the linear fits that converge to $T_c$ at $L^{-1/\nu}=0$ for conventional \textit{WLS}
and the new procedure, both using the $80\%$ flatness criterion. The final estimate for $T_c$ was taken as the
mean value obtained from both fits.

Since $T_c$ is now estimated, we can calculate the critical exponents $\beta$ and $\gamma$. According to Eq.\eqref{exp2},
the maximum of the finite-lattice susceptibility defined by Eq.\eqref{ki} is asymptotically proportional to $L^{\gamma/\nu}$.
In Fig.\ref{gamma} we show these results for the conventional \textit{WLS} and our procedure, both using the $80\%$-criterion
of flatness. In the vicinity of the critical temperature the magnetization scales as $L^{-\beta/\nu}$.
We can use Eq.\eqref{exp1} at the critical point to calculate the exponent $\beta$ directly from the slope of the log-log
graph and find $\beta$. In Fig.\ref{beta} we show again the results for conventional \textit{WLS} and our procedure for this
exponent. One can see that in all cases our procedure is more accurate than the conventional \textit{WLS}.

For the conventional \textit{WLS} and the new procedure proposed here, simulations were carried out using
$80\%$ and $90\%$ flatness criterions and for \textit{1/t} scheme the simulations were halted for $\ln f = 10^{-6}, 5.10^{-7},
10^{-7}, 5.10^{-8}$ and $10^{-8}$. In Table \ref{table} we show the results for the $1/t$ simulations, the conventional
\textit{WLS} and our procedure along with the exact values. The $1/t$ results become accurate only when $\ln f\sim 5\times10^{-8}$,
and for lower values ​​of $\ln f$ they become worse, giving the impression that they are already fluctuating around the true value.
The conventional \textit{WLS}, displays problems of accuracy, while our results are adequately accurate for both 80\% and 90\%
flatness criterions. It is worthwhile mentioning that we have obtained high-resolution values using the 90\% flatness
criterion, which should be compared with the erratic behavior of the $1/t$ simulations for $\ln f<5\times10^{-8}$, but
such stringent level of flatness is difficult to apply to other systems \cite{blume,bjp,cpc,seaton}, resulting sometimes in
non-convergence or even more inaccurate values. Moreover the 90\% flatness criterion simulations are very time consuming.
We conclude therefore that the widely adopted 80\% flatness criterion is indeed the best guess, since it is applicable to all systems.

\maketitle

\section{Application to a self-avoiding homopolymer}

In this section, we apply to a homopolymer the initial tests to determine up to which modification factor one should continue
the $WLS$ and from which $f_{micro}$ the microcanonical averages should be accumulated. We consider a homopolymer consisting of
$N$ monomers which may assume any self avoiding walk (SAW) configuration on a two-dimenstional lattice.

\begin{figure}[!ht]\centering
\begin{center}
 \includegraphics[width=.9\linewidth]{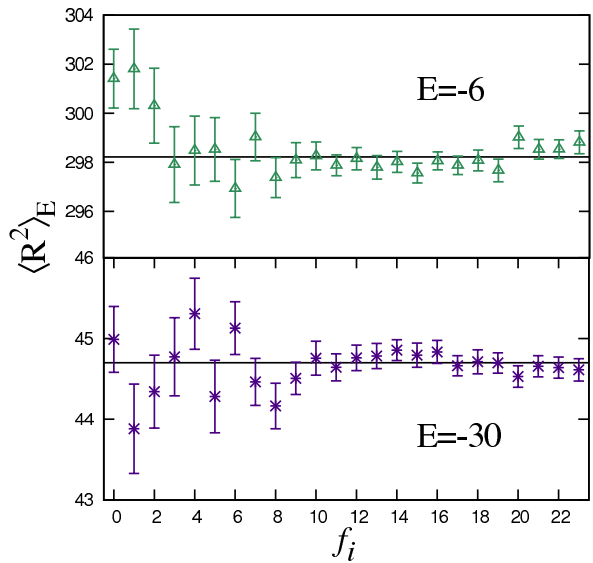}
\end{center}
\caption{Evolution of \hspace{.1cm} the \hspace{.1cm} microcanonical average of the mean square end-to-end distance for $N=50$ at $E=-6$ and $E=-30$
during the simulations over $100$ independent runs for each flatness stage.}
\label{micro_pol}
\end{figure}

Assuming that the polymer is in a bad solvent, there is an effective monomer-monomer attraction in addition to the
self-avoidance constraint representing excluded volume. For every pair of non-bonded nearest-neighbor monomers the
energy of the polymer is reduced by $\varepsilon$. The Hamiltonian for the model can be written as
\begin{equation}
\mathcal H=-\varepsilon\sum_{<i,j>}{\sigma_i\sigma_j},
\end{equation}
where $\sigma=1(0)$ if the site $i$ is occupied(vacant), and the sum is over nearest-neighbor pairs\cite{dickman}. (The
sum is understood to exclude pairs of bonded segments along the chain.) We used the so-called reptation or
\textit{``slithering snake''} move which consists of randomly adding a monomer from one end of the chain and removing
a monomer from the other end, mantaining the size of the polymer constant.We define one Monte Carlo step as $N$ attempted
moves.

In this model, besides the energy, another quantity of interest is the mean square end-to-end distance given by
\begin{equation}
\langle R^2\rangle=\langle[(x_N-x_0)^2+(y_N-y_0)^2]\rangle
\end{equation}

\begin{figure}[!ht]\centering
\begin{center}
 \includegraphics[width=.7\linewidth,angle=-90]{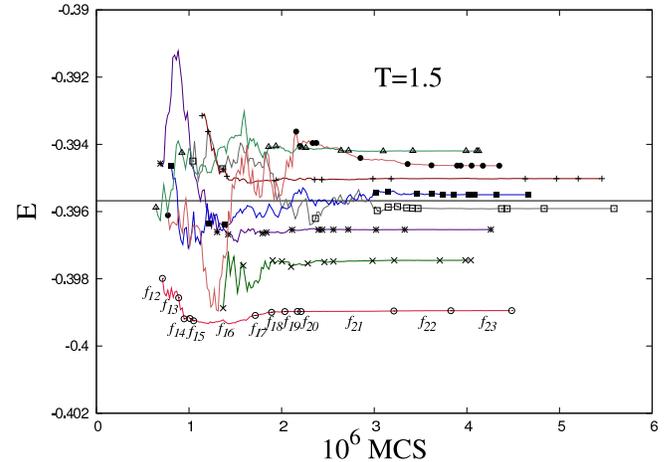}
\end{center}
\caption{Evolution of the energy at $T=1.5$ during the $WLS$ for eight independent runs beginning from $f_{12}$.
The straight line is the mean value obtained from $100$ independent runs.}
\label{medpol}
\end{figure}
\begin{figure}[!ht]\centering
\begin{center}
 \includegraphics[width=.7\linewidth,angle=-90]{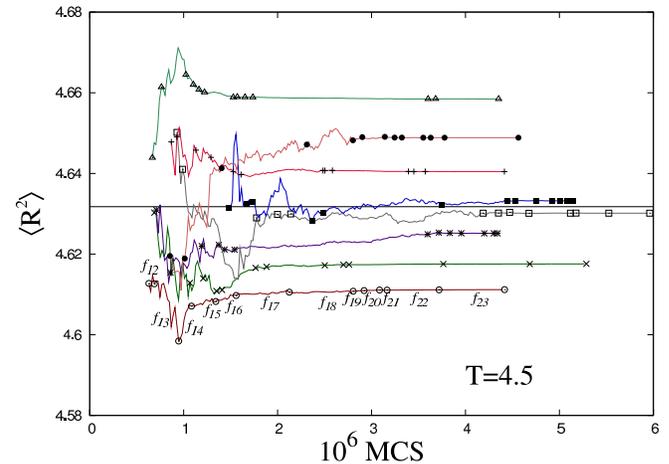}
\end{center}
\caption{Evolution of the mean square end-t-end distance at $T=4.5$ during the $WLS$ for eight independent runs
beginning from $f_{12}$. The straight line is the mean value obtained from $100$ independent runs.}
\label{end_to_end}
\end{figure}
As prescribed in Section II, we updated the density of states and the histogram only after each Monte Carlo step.
In order define from which modification factor to begin accumulating the microcanonical averages we estimated
$\langle R^2\rangle_E$ for each $f_i$ during the simulations for several energy levels. In Fig.\ref{micro_pol}
we show these results for $E=-6$ and $E=-30$ (the ground state of the $N=50$ homopolymer is $E_{min}=-36$). One
can see that the microcanonical averages should be accumulated from $f_{micro}=f_{10}$. On the other hand, to
estimate $f_{final}$ for halting the simulations, we calculated the canonical average of the energy and the mean square
end-to-end distance during the simulations for five fixed temperatures, namely, $T=0.5, 1.5, 2.5, 3.5$ and $4.5$. In Fig.\ref{medpol} we show the behavior of the energy at $T=1.5$ and in Fig.\ref{end_to_end} the behavior of the mean square
end-to-end distance at $T=4.5$. All the graphs we have constructed for these quantities for the temperatures mentioned above have
similar performance. We see therefore that for this model the simulations should be carried out up to $f_{final}=f_{18}$.

Other more elaborate models, such as the HP model of protein folding \cite{hp} or continuous (off-lattice) models of polymers
\cite{wust,binder} will require a test similar to the one made ​​in this section that may point to different values ​​for $f_{micro}$ and $f_{final}$. Notwithstanding that, our results suggest that this final modification factor will occur far before $f_{26}\approx1+10^{-8}$, leading to considerable CPU time savings.

\section{Conclusions}

We have demonstrated that the conventional \textit{WLS} presents problems of accuracy, but with very few changes in the
implementation of the method, namely, updating the density of states only after each Monte Carlo step, halting the simulations
for $\ln f<\ln f_{final}$, with $f_{final}$ determined by the canonical averages during the simulations and accumulating the microcanonical averages only for $\ln f<\ln f_{micro}$, where $f_{micro}$ is found from the behavior of the microcanonical
averages in each modification factor, it becomes quite accurate. Adopting the Monte Carlo step to update the density of
states and delaying the start of the microcanonical averaging are changes that lead to a improved accuracy of the algorithm,
while the proper definition of when to stop the simulation ($f_{final}$) saves a lot of CPU time.

It should be pointed out that a direct comparison of
the density of states with exact calculations, although pictorially very impressive, is not a good test for algorithms
that estimate the density of states. The canonical and microcanonical averages during the simulations are a more adequate
checking parameter for convergence. Another important conclusion is that no single simulation in particular tends to the exact value. One can obtain results as close as possible to the exact value by increasing the number of independent runs.

The great advantage of our findings is that all existing codes using \textit{WLS} can be promptly adapted to this
new procedure just adding a few lines to the computer program.
\vspace{.5cm}
\section{Acknowledgment}

This work was supported by Brazilian agencies CNPq and FUNAPE-UFG.

We thank Salviano de Araújo Leão for his helpful and substantial advising and support in computational abilities.


\begin{thebibliography}{99}

\bibitem{wls1} F. Wang, D. P. Landau, Phys. Rev. Lett. \textbf{86}, 2050 (2001).

\bibitem{wls2} F. Wang and D. P. Landau, Phys. Rev. E. \textbf{64}, 056101 (2001).

\bibitem{zhou} C. Zhou and J. Su , Phys. Rev. E \textbf{78}, 046705 (2008).

\bibitem{belardinelli} R. E. Belardinelli and V. D. Pereyra, Phys. Rev. E \textbf{75}, 046701 (2007).

\bibitem{belardinelli2} R. E. Belardinelli, S. Manzi, and V. D. Pereyra, Phys. Rev. E \textbf{78}, 067701 (2008).

\bibitem{belardinelli3} R. E. Belardinelli and V. D. Pereyra, J. Chem. Phys. \textbf{127}, 184105 (2007).

\bibitem{plichske} See, e.g., M. Plichske, B. Bergersen, Equilibrium Statistical Physics,
Cambridge University Press, 2008.

\bibitem{swetnam} A. D. Swetnam and M. P. Allen, J. Comput. Chem. n/a. doi: 10.1002/jcc.21660.

\bibitem{troyer} P. Dayal, S. Trebst, S. Wessel, D. Wurtz, M. Troyer, S. Sabhapandit, and
S. N. Coppersmith, Phys. Rev. Lett. 92, 097201 (2004).

\bibitem{bhatt} C. Zhou and R. N. Bhatt, Phys. Rev. E 72, 025701(R) (2005).

\bibitem{beale} P. D. Beale, Phys. Rev. Lett. \textbf{76}, 78 (1996).

\bibitem{mcs} A Monte Carlo sweep consists of $L^2$ spin-flip trials.

\bibitem{fisher1} M. E. Fisher, in \textit{Critical Phenomena}, edited by M. S. Green (Academic, New York, 1971).

\bibitem{fisher2} M. E. Fisher and M. N. Barber, Phys. Rev. Lett. \textbf{28}, 1516 (1972).

\bibitem{barber} in \textit{Phase Transitions and Critical Phenomena}, edited by C. Domb and J. L. Lebowitz
(Academic, New York, 1974), Vol. 8.

\bibitem{blume_l2} The study performed in \cite{blume} was carried out updating the density of states after every $L^2$
trial moves. Thanks to a fortunate misunderstanding, the authors have adopted the Monte Carlo sweep by analogy with the Metropolis
algorithm.

\bibitem{blume} C. J. Silva, A. A. Caparica and J. A. Plascak, Phys. Rev. E \textbf{73}, 036702 (2006).

\bibitem{bjp} C. J. Silva, A. G. Cunha-Netto, A. A. Caparica and R. Dickman, Braz. J. Phys. \textbf{36}, \textit{no. 3A} 619 (2006).

\bibitem{cpc} A. G. Cunha-Netto, R. Dickman and A. A. Caparica, Comput. Phys. Comm. \textbf{180}, 583 (2009).

\bibitem{seaton} D. T. Seaton, T. W\"{u}st, and D. P. Landau, Phys. Rev. E, 81, 011802 (2010).

\bibitem{dickman} R. Dickman, J. Chem. Phys. \textbf{96}, 1516 (1992).

\bibitem{hp} T. W\"{u}st and D. P. Landau, Comput.Phys.Comm. \textbf{179}, 124 (2008).

\bibitem{wust} T. W\"{u}st and D. P. Landau, Phys. Rev. Lett. \textbf{102}, 178101 (2009).

\bibitem{binder} M. P. Taylor, W. Paul and K. Binder, J. Chem. Phys. \textbf{131}, 114907 (2009).

\end{thebibliography}
\end{document}